\documentclass{article}
\usepackage{amsmath}
\usepackage{graphicx}

\title{Defining the Entropy and Internal Energy of a Monetary Schelling model through the Energy States of Individual Agents}

\author{
George-Rafael Domenikos\thanks{Laboratory of Applied Thermodynamics, Mechanical Engineering Department, National Technical University of Athens, rdomenikos@mail.ntua.gr},
Tyler Laurie\thanks{Department of Statistics and Data Science, University of Central Florida; tlaurie@knights.ucf.edu},
Sahar Awaji\thanks{Department of Statistics and Data Science, University of Central Florida; sa.awaji2014@knights.ucf.edu},
Alexander V. Mantzaris\thanks{Department of Statistics and Data Science, University of Central Florida; alexander.mantzaris@ucf.edu}
}

\begin{document}

\maketitle
\begin{abstract}
This work investigates a modified Schelling model within the scope and aims of Social Physics. The main purpose is to see if how the concepts of potential and kinetic energy can be represented within a computational sociological system. A monetary value is assigned to all the agents in the Monetary Schelling model and a set of dynamics for how the money is spent upon agent position changes and gradual loss. The introduction of the potential and kinetic energy allows for the entropy to be calculated based upon the distribution of the agent energies and as well as the internal energy of the system at each time point. The results show how the movements of the agents produce identity satisfactions with their neighbors decreasing the internal energy of the system along with the decay in the monetary holdings. Simulations are run where agents are provided monetary values at fixed intervals and this causes a subset of the agents to mobilize and explore new positions for satisfaction and increases the entropy with the internal energy removing the system from the fixed point.
\end{abstract}


\section{Introduction}

Socio physics is a field with the goal of understanding social processes and phenomena using methodological approaches derived or inspired by the science of physics \cite{stewart1948concerning,stewart1950development}. This goal does not aim to ignore the work done or ongoing from social sciences but to augment it with different tools to help dissect the granularity of the intricate processes under study. Such processes include political polarization \cite{gentzkow2016polarization}, policy making \cite{harbridge2014public}, morality consensus development \cite{cohen1993moral} and many others. These processes pose interesting challenges to researchers as they display non-linear phenomena which are a challenge to model as noted in the seminal work of \cite{chen1991self,bak1988self} studying self-organized criticality. Physics has developed many approaches to studying complex processes which accurately predict their behaviors such as that of transistors \cite{miller1992physics} where high degrees of confidence are applied to the models. It is more important that these processes are understood as urbanization is on the increase globally \cite{cohen2006urbanization}. 

The work presented here proposes novel methodological advancements in Socio Physics using a modification of the well known Schelling model of segregation \cite{schelling1971dynamic,schelling2006micromotives}. When Schelling introduced this model it involved a 13x13 grid where each grid cell could be occupied by at most a single hypothetical agent. There were empty cells and each agent had one of 2 different identity labels. Agents contain a binary state variable of identity satisfaction whose value is defined by having a sufficient number of adjacent cells occupied by agents of the same identity label. If the threshold for the number of homogeneous neighboring agents is present the agent \emph{remains} in the same position, or else it will move to another position. This is the fundamental dynamic of the Schelling model and is evaluated for all the agents by choosing agents in a random order at each application of the dynamic. Modifications and generalizations of this canonical Schelling model exist \cite{rogers2011unified} producing interesting behaviors.

The Schelling model is one of the first sociological models explored in a computational environment as Schelling conducted the simulations by hand originally. Similarly the Schelling model is one of the first to be explored as having dynamics that are analogues of physical processes. A parallel with the Ising model of magnetism \cite{ising1925contribution,brush1967history,cipra1987introduction} was seen as having a similar dynamic since the atomic units in the Ising model also take a summation of their adjacent neighbor states of spins rather than identities. The fundamental connection between this sociological model and physical model is explored in \cite{stauffer2007ising}. The canonical model provides a framework which can be used to explore a myriad of phenomena such as in \cite{muller2008inhomogeneous} that investigates the self-organized temperature of a Schelling dynamic with Ising model descriptions. 

An important practical study using the Schelling model, although void of physical analogues, is that of \cite{hatna2012schelling,hatna2014combining} which from empirical observation found that the wealth of citizens plays a large role in urban mobilizations. Using these findings the work of \cite{mantzaris2020incorporating} proposes a dual dynamic Schelling model that uses the identity dynamic and in parallel a monetary dynamic upon a financial store held by the agents in which the entropy trace of the model can be explored. In contrast to the findings that the canonical Schelling model has a decreasing entropy trace as more agents enter the remain state (\cite{mantzaris2018examining}), the results of \cite{mantzaris2020incorporating} show that the entropy trace increases overall due to the monetary dynamics therefore correcting the decreasing entropy trace. In \cite{mantzaris2018examining,mantzaris2020incorporating} the entropy is computed using the Boltzman statistics where the distribution of the macrostates is found via microstate sampling. This involves a considerable number of Monte Carlo samples to capture the mode density and surrounding regions. 

Important recent work in this field includes \cite{papanikolaou2022consensus} that discusses a fundamental modeling paradigm to the formation of consensus. In terms of socio-physics it provides novel insight on physical concepts like the magnetization can be incorporated into a model of social processes. In the work of \cite{schweitzer2022social} the reader can find fundamental developments for the field of socio-physics in that the concepts of percolation and nucleation introduced into a simulated model of social interactions. Percolation is an important concept as the paper describes since the thresholds allow for phase transitions to be modeled with it evident in social processes which helps drive the investigation for how an understanding of physics can help model even those systems. The work of \cite{metz2022policy} provides a deep investigation into policy networks for a subset of European countries and how power can concentrate which is relevant to the modeling paradigm utilized in this study. A key point in the paper is the formation of ties and the association with a concentration of power when a complex network is produced. This also sets of a foundation for the definition of variables which will be treated thermodynamically in this work and the future. This work based on the Schelling model will utilize the concept of the 'agent' for each resident and as a system it is an agent based system (ABM \cite{epstein2014agent_zero}). The work of \cite{herzog2023nonlinear} shows the nonlinear behaviors of agents' collective behavior in relation to polarization and how institutions can affect the trajectory of such trends.

The goal of this paper within the context of socio physics is three fold. First to present a Monetary Schelling model for which the energy, kinetic, and potential, are computed for each agent at each iteration (time step). Second, calculate the entropy of the model based upon the distribution of the energies of the agents (sum of kinetic and potential energy) after discretization of the energy values. Such an approach avoids the costly Monte Carlo sampling scheme which was previously used. Third to calculate the internal energy of the model that is based upon the energies of the agents and the probability of their presence. The introduction of these quantities into the model will allow for a deeper physical interpretation of the Schelling model and dynamics as well as exploring the intricacies of how the monetary and identity dynamics cooperate in order to produce an overall stable system dynamic. Some modifications to the canonical Schelling is that agents upon movement subtract a portion of their monetary value and distribute it uniformly to the agents in the new neighborhood (as in \cite{mantzaris2020incorporating}). Agents at each iteration lose a percentage of their monetary value which is not absorbed by other agents. In order to move positions agent must have a larger kinetic energy than a minimum threshold.

Although this work relies on the basic Schelling construction where agents move on grid (lattice) the work of \cite{zingg2019entropy} develops a methodology which allows the entropy for a network to be computed. These networks do not follow the rigid uniform pattern offered by a grid and can apply directly to complex real situations.  In relation to previous work this work offers the exploration of internal energy and kinetic energy of the system. For a model such as the Schelling model (Monetary Schelling model in this case) this has not been explored yet.

The Methodology section will present the definitions of the dynamics and the Results the simulation trajectories of the quantities represented of the model in the definitions. A key finding is that entropy trace results conform with the results of \cite{mantzaris2018examining} and that internal energy of the model follows a trajectory of decay. The simulations also explore the situation where the agents are provided a uniform monetary addition every 200 iterations. This injection allows the model to deviate from the fixed resting point and re-establish the previous dynamics providing insight into how financial policy can affect urban mobilizations.
The implementation has been done using Julia Lang \cite{bezanson2012julia} as it offers computational efficiency and clear syntax for the representation of the equations defined.

\section{Methodology}
\label{sec:Methodology}

This section defines the quantities and values which parameterize the dynamics of the Monetary Schelling model explored in this work. Key aspects which deviate from the canonical Schelling model \cite{schelling1971dynamic} are that agents contain a monetary store which can increase or decrease through environment injections, from movements that induce costs, loss over time to the environment and spreading some expenses to the neighborhood. On top of the monetary quantities and introduced dynamics physical quantities are defined based on the state of the agents at each time point: kinetic and potential energy. This allows for the entropy of the system at each time point to be calculated using the distribution of the energies among the agents in the same manner as it is done in kinetic theory of gases. Using the discretized probability distribution of the agent energies the internal energy of the system is then be calculated. 

The potential energy $V$ for each agent at position $(i,j)$ at simulation time point $t$ can be found using the state of agents' \emph{remain} status $r_{i,j,t} \in [0,1]$:  
\begin{equation}
\label{eq:PotentialEnergy} 
  V_{i,j,t} =
  \begin{cases}
    0 & \text{if } r_{i,j,t-1} = 1, \\
    1 & \text{if } r_{i,j,t-1} = 0.
  \end{cases}
\end{equation}
This represents the intuition that if an agent will remain in the same position between time steps then the potential energy is zero as it is not mobile. As will be seen this remain state $r_{i,j,t}$ is not only dependent upon the local neighbor homogeneity count as in the classic canonical Schelling model, but also upon there being a sufficient kinetic energy value. Therefore the remain $r_{i,j,t}$ will depend upon a both the local identity differential and minimum monetary value (kinetic energy expenditure threshold). 

The kinetic energy value for each agent at each time point $(i,j,t)$ based upon the value of the monetary store it holds, $m_{i,j,t}$:
\begin{equation}
\label{eq:KineticEnergy} 
    K_{i,j,t} = \frac{2m_{i,j,t}}{\text{max}(\mathbf{m}_{i,j,t=0}: \forall i,j \notin \mathbf{m}_{i,j} = \emptyset )}.
\end{equation}
Here, $\mathbf{m}$ represents the vector of all the agent monetary store values. The denominator of this equation finds the maximum monetary among all the agents value which exists at the start of simulation. This type of normalization is done in order to have comparable magnitudes between the kinetic energy and the potential energy. As will be seen agent movements will spread the movement costs around to their neighbors causing a decrease in the maximum monetary value held by agents at later time points resulting in $K_{i,j,t}$ being typically less than 1. Another aspect is that the kinetic energy although dependent upon a constant and variable state for its value the normalization aspect is not a function over time. The value of $K_{i,j} \in [0,2]$ due to the constant term 2 included in the numerator which is chosen to facilitate the movements of the agents. If this factor was not included then the kinetic energy would not surpass the value of 1 which is what the potential energy value can have. The kinetic energy value must be at times greater than the potential energy in order to ensure movements. A larger value for this can be selected which would promote a greater amount of freedom of movement to the agents and the modeler can change this value if needed.

The classic canonical Schelling model agent identity homogeneity satisfaction criteria is represented by $l(i,j,t) > h_{i,j}$. The number of local homogeneous agents neighboring the position $(i,j)$ at time point $t$ is given by $l(i,j,t)$ and the threshold for that grid position is given by $h_{i,j}$. The $h$ value does depend upon the grid position since the corners and edges of the grid require fewer homogeneous neighbors (as implemented in the code running the simulations). This proposed model introduces a quantity, $\Delta \epsilon = 0.2$, that is a minimum kinetic energy cost required by agents in order to change grid positions. In this Monetary Schelling model the remain state for an agent over time, $r_{i,j,t}$ is given by:
\begin{equation} 
    \label{eq:Remain}
    r_{i,j,t} = ( K_{i,j,t} > \Delta \epsilon ) \land (h(i,j,t) < h).
\end{equation}
This states that whether an agent remains in the same position or moves is dependent upon both the local identity homogeneity and whether the agent contains enough kinetic energy to surpass the barrier cost (akin to the activation energy, or potential barrier) $\Delta \epsilon = 0.2$. In the implementation agents move to a random grid position when they do not remain in their current position rather than scan the grid to where there is an identity satisfaction to be arrived at in order to be closer to a physical system without guidance on the particles. 

Agents are exposed to a dynamic of a movement cost which affects their monetary store value. When an agent moves its monetary value stored, and consequently its kinetic energy, will be reduced with each movement. This reduction in the monetary value is the inverse of the kinetic energy cost:
\begin{equation}
    \label{eq:MovementMoneySpending}
    m_{i,j,t} = m_{i,j,t-1} - K^{-1}(\Delta v).
\end{equation}
This dynamic based on the value $\Delta v$ can be seen as type of monetary quanta. This reduction in monetary value is then uniformly distributed amongst its neighbors which then have their monetary value (and kinetic energy) increased. Agents also undergo a monetary decrease on each iteration independently of other agents as an analogue to the expenses residents are expected to have over time other than just costs related to relocation. This is modeled by having the monetary holding of each agent reduced by 5\% and this monetary loss is not given to other neighbors but simply subtracted from the system (akin to heat radiation). The $K^{-1}(\Delta v)$ is the inverse function of the value of $\Delta v$ so that what is returned is the monetary value that corresponds to smallest amount of kinetic energy which can be exchanged. In this approach the model introduces the concept of a 'monetary quanta' which is to be the smallest amount of money that can be exchanged as an analogue to the smallest amount of kinetic energy that can be exchanged as in a physical system. In terms of the simulated reality the hypothetical residential agents experienced this corresponds to a minimal transaction fee for all transactions.

The energy of each agent at every time point in the simulation is defined by the aggregate of the potential and kinetic energy of the agents:
\begin{equation}
\label{eq:TotalEnergy}
    E_{i,j,t} = V_{i,j,t} + K_{i,j,t}.
\end{equation}
This represents the total energy of each agent during the course of the simulation.

The probability of the energy states is required in order to calculate the entropy and internal energy of the complete system at each time point. Since the energies are on a continuous domain, they are discretized first. Each energy state value is denoted with $n$ and $n_{total}$ as the total number of discrete energy states of all the agents at each time point independently. The discretization uses fixed energy bins set at, $E_n \in \{0, 0.2, 0.4, \ldots, 999.8, 1000\}$ and a bin distance threshold of $\epsilon = 0.1$:
\begin{equation}
    \label{eq:EnergyProbability}
    p_t(E_n) = \frac{\sum_{i=1}^N \sum_{j=1}^N  (|E_{i,j,t} - E_{n}| \leq \epsilon \land E_{i,j,t} \ne \emptyset)}{\|N\|}.
\end{equation}
Here $\|N\|$ is the total number of agents in the system which remains constant. $p_t(E_n)$ is a probability mass function over the discrete domain of possible energy states $E_n$. 

With the probabilities of the energy state values at each time point the entropy can be calculated using distribution of the probabilities with the standard formulation \cite{shannon1948mathematical}:
\begin{equation}
    \label{eq:ModelEntropy}
    S_t = - \sum_{n=1}^{n_{total}} p_t(E_n) ln(p_t(E_n)).
\end{equation}
The interpretation of the value of $S_t$ can be understood as a measure of the uniformity of the energy values across the domain, so that larger $S_t$ values are a result of the agents occupying a broad range of $E_n$ values and a low $S_t$ when the agents occupy a limited number of $E_n$ values. This concept of how entropy relates to spatial configurations of agents on a grid can be understood thoroughly in the detailed work of \cite{styer2000insight}. Over the simulation iterations as the system converges towards homogeneity the entropy will be decreasing and increasing if the agents begin to occupy a wide range of energy states. $S_t$ can then provide insight into the stability of the system by assessing the range of different energy states the society residents occupy.
This approach bypasses the need to produce a lengthy Monte Carlo estimate of numerous independent microstate samples in order to find a distribution on the macrostates as is done in \cite{mantzaris2018examining,mantzaris2020incorporating} for Schelling based models, and in \cite{domenikos2022model,mantzaris2023exploring} for non-Schelling computation sociological models.

The internal energy of the system at each time point is defined by using the distribution of the energies and the energy values themselves:
\begin{equation}
    \label{eq:ModelInternalEnergy}
    U_t = \sum_{n=1}^{n_{total}} p_t(E_n) E_n.
\end{equation}
This quantity helps understand if the system will be energetic in the future or not. Systems with zero internal are at a stable equilibrium or resting point.

\section{Results}

\subsection{Initial Investigation}
\label{subsec:InitialResults}

\begin{figure}[h]
\begin{center}
    \includegraphics[width=1.0\textwidth]{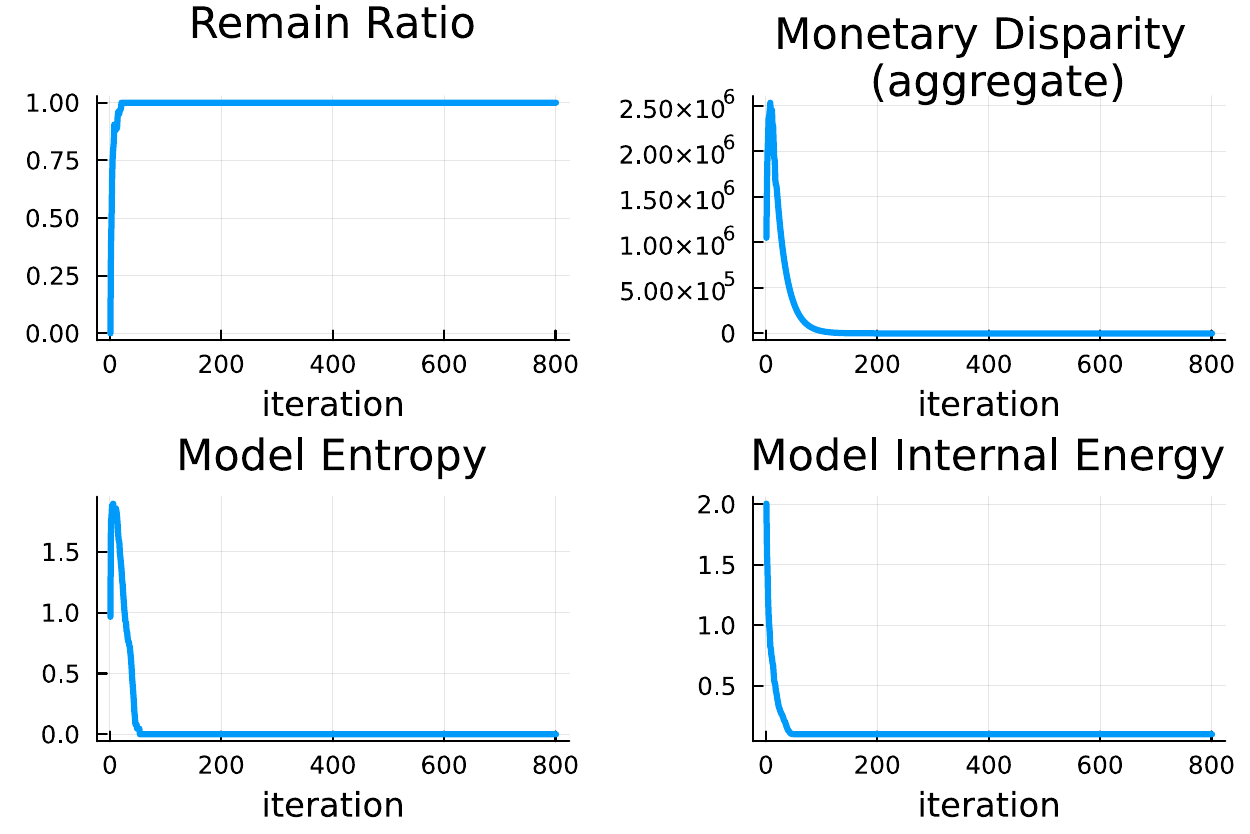}    
\end{center}
\caption{The horizontal axis on each plot shows the iteration of the Monetary Schelling model. Top-left: the percentage of agents which remain in their position due to Schelling identity satisfaction. Top-right: the overall aggregate monetary disparity between each agent and their neighbors. Bottom-left: the model entropy calculated from the probability distribution of the agent energies (Eq~\ref{eq:ModelEntropy}). Bottom-right: the internal energy of the system based upon the probability of an energy and that energy value (Eq~\ref{eq:ModelInternalEnergy}). The system arrives to rest and stabilizes around a fixed point as agents remain in their positions.  \label{fig:InitialResults}}
\end{figure}

Figure~\ref{fig:InitialResults} shows four plots of a simulation of the Monetary Schelling model for 800 iterations (iterations marked on the horizontal axis). The plot on the top left shows the percentage of agents at each iteration which will 'remain' (Eq~\ref{eq:Remain}) in their grid position due to the criteria of the Monetary Schelling model discussed in the Methodology. It must be noted that in this model although all the agents converged to a remain state in their current position the reason may be due to Schelling identity satisfaction or to an insufficient amount of monetary holdings that provide the necessary kinetic energy $\Delta v$ required to change positions. It can be seen how even in this model almost all the agents arrive to be in a remain state quickly from the random initialization as noted originally by Thomas Schelling \cite{schelling1971dynamic}. The top right plot shows the value of the overall monetary disparity accumulated across all the agents in the model. This measure takes the absolute monetary value difference between an agent and all adjacent neighbors without any normalization to resemble physical energetic calculations. The bottom left plot shows how the model entropy for the Monetary Schelling model changes value with the new proposed methodology (Eq~\ref{eq:ModelEntropy}). The eventual decrease in entropy confirms the findings in \cite{mantzaris2018examining} where the entropy of the system decreases as the system arrives at a more 'organized' agent identity configuration. The bottom right plot shows how the internal energy (Eq~\ref{eq:ModelInternalEnergy}) of the Monetary Schelling model can be monitored. The monotonic decrease is due to agents having less potential energy being in an identity satisfaction (less potential energy) and reduced amount of monetary holding (less kinetic energy). Monetary holdings are driven to zero in this simulation as the agents will spend a portion of their holdings regardless of movement and there is no dynamic for re-introducing monetary values to the agents. This shows how the model arrives at a static equilibrium similar to physical systems since there is no energy in the system participants (agents). 

\begin{figure}[h]
\begin{center}
    \includegraphics[width=1\textwidth]{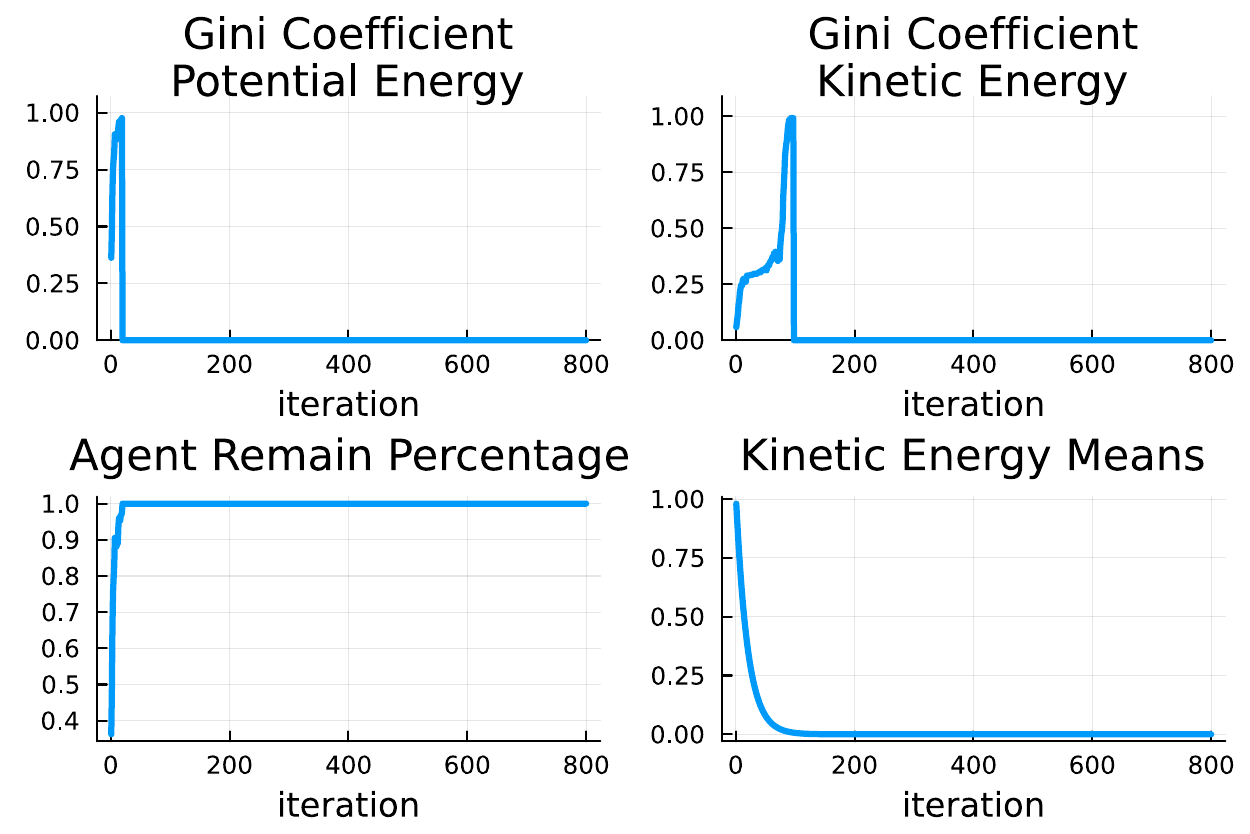}
\end{center}
\caption{The top plots show the Gini coefficient values for the potential energies of the agents (Eq~\ref{eq:PotentialEnergy}) (based upon their remain values) and the kinetic energies (Eq~\ref{eq:KineticEnergy}) (based on the amount of money held by each agent). The bottom left plot shows the percentage of agents which are in the remain state during the simulation and the bottom right plot the kinetic energy mean at each iteration. This shows that there is a phase where most agents no longer change positions and gradually lose their kinetic energy over time due to holding less monetary value. \label{fig:InitialResultsGini}}
\end{figure}

Figure~\ref{fig:InitialResultsGini} shows a different set of analytic measurements of the same system simulation presented in Figure\ref{fig:InitialResults}. The top left plot shows the value of the Gini coefficient computed on the value of the potential energy (Eq~\ref{eq:PotentialEnergy}) at each time point. The increase continues until there is only a single agent which has an opportunity to move (maximum energetic inequality). This can come from an agent not being identity satisfied but they must also have the sufficient monetary values to move (Eq~\ref{eq:Remain}). The top right plot shows the Gini coefficient for the kinetic energy (Eq~\ref{eq:KineticEnergy}) which is not tied to the potential energy (Eq~\ref{eq:PotentialEnergy}) as it decreases at a later iteration. The decrease happens because of the spending at each iteration regardless of movements. It continues to grow until only a single agent has a non-zero equivalent monetary value. The bottom left plot shows the percentage of agents which will remain in their positions in the next iteration. The bottom right plot is the mean kinetic energy of the agents. 

\subsection{Monetary Injections}
\label{subsec:MonetaryInjections}

The results of the simulations shown in this subsection demonstrate the effects of 'injecting' monetary values into each of the agents uniformly at every 200 iterations (100,000 to each agent). The purpose is to re-stimulate the system from the fixed point of stable equilibrium (resting point) and observe the trends which take effect.

\begin{figure}[h]
\begin{center}
    \includegraphics[width=1.0\textwidth]{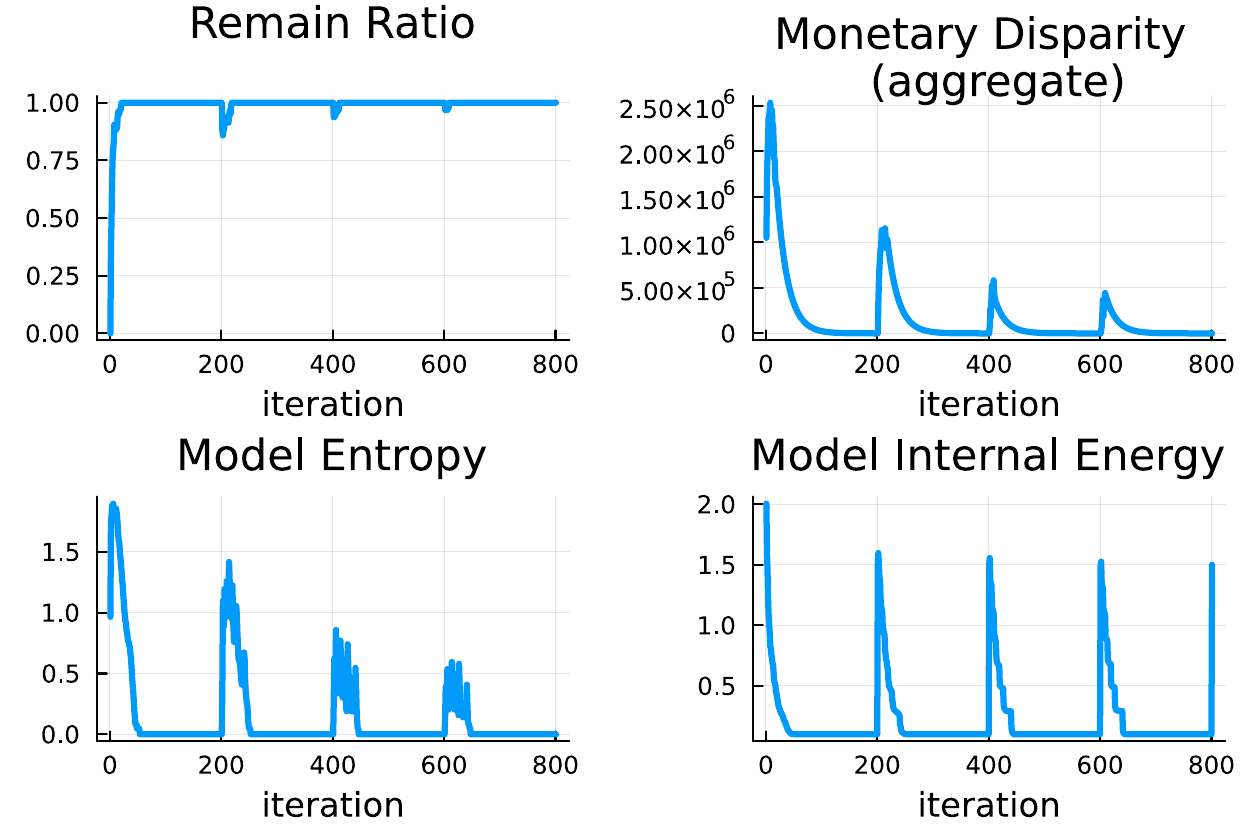}    
\end{center}
\caption{The horizontal axis on each plot shows the iteration of the Monetary Schelling model. The top-left plot shows the percentage of agents which remain in their position due to Schelling identity satisfaction. Top-right is the overall aggregate monetary disparity between each agent and their neighbors. Bottom-left is the model entropy trajectory (Eq~\ref{eq:ModelEntropy}) which is seen to reflect the size of the set of the mobile agents over time. Bottom-right shows the internal energy of the system (Eq~\ref{eq:ModelInternalEnergy}) based upon the probability of an energy and the energy value which is also related to the number of agents which are mobilized. \label{fig:InjectionResults}}
\end{figure}

Figure~\ref{fig:InjectionResults} shows in the top left plot the remain ratio which does not display monotonicity with the monetary injections. The configurations of the agents does not allow all the agents to find identity satisfaction and previously would stop moving (remaining) due to insufficient funds. With the injections the unsatisfied subset of agents begins to mobilize again as their kinetic energy is sufficient to overcome the $\Delta v$ (Eq~\ref{eq:Remain}). As each injection is made a fewer amount of agents still mobilize some eventually find new homogeneous clusters among their group with each injection (\cite{domenikos2022model}). The top right plot shows the aggregate monetary disparity between the agents and their adjacent neighbors. It can be seen how the injections cause a monetary disparity since agents which still move will have less monetary values  between neighbors creating the disparity. This relative size of these spikes mirrors the dip sizes in the top left plot so that they decrease as the number of mobile agents decreases as well from there being a smaller set of identity unsatisfied. The bottom left plot shows how the entropy of the model is affected by the monetary injections. The entropy has a decreasing trend due to the distribution of the total agent energy probabilities accumulating into a single discrete bin. The injections allowing for a portion of agents to deviate from the uniformity causes the distribution to have multiple probability bins and therefore an entropy. The model internal energy shows a similar spiking but the size of the spikes are not dependent upon a distributional disparity but only upon the energy of the agents which is uniform at each injection. The initial internal energy value is highest since the potential energy among the agents is highest even if the monetary injections covered decreases in the kinetic energy.

\begin{figure}[h]
\begin{center}
    \includegraphics[width=1\textwidth]{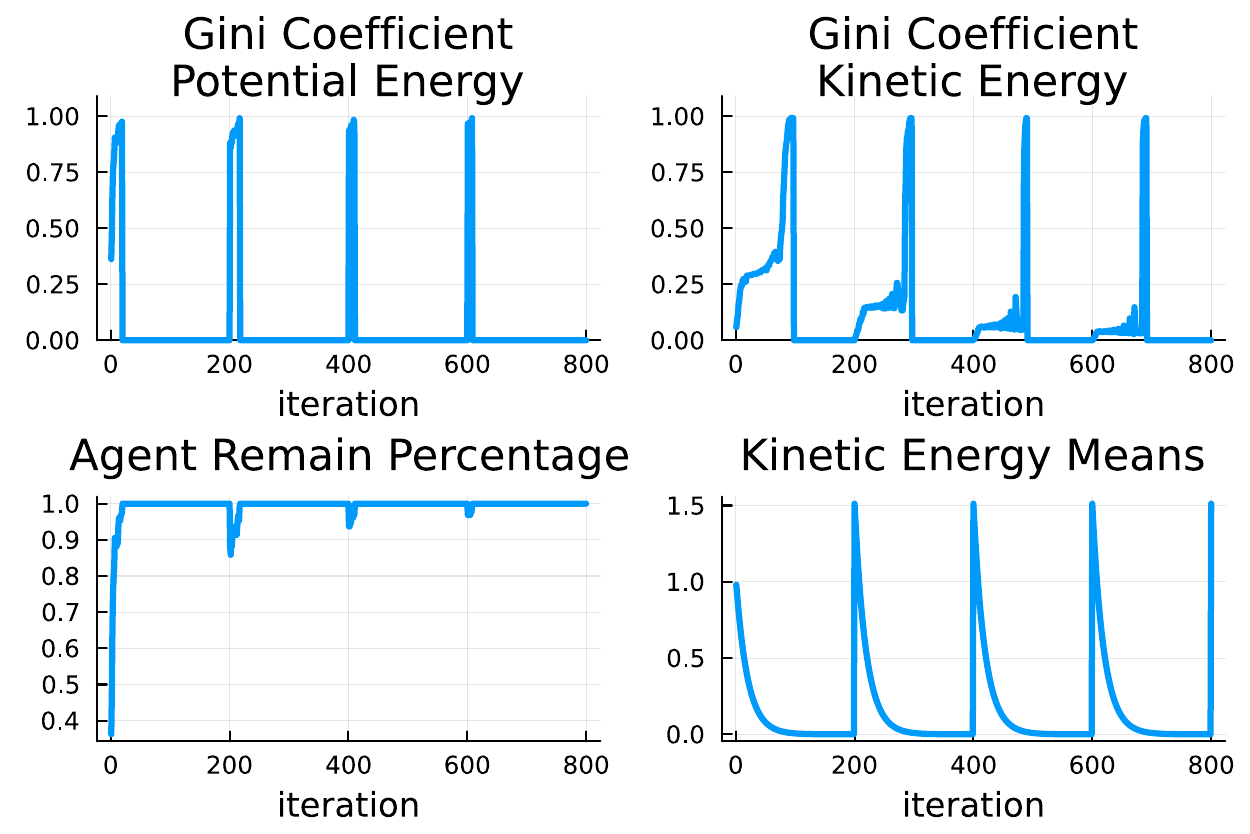}
\end{center}
\caption{The top plots show the Gini coefficient values for the potential energies of the agents (based upon their remain values) Eq~\ref{eq:PotentialEnergy}, and the Gini of kinetic energies (based on the amount of money held by each agent Eq~\ref{eq:KineticEnergy}). The bottom left plot shows the percentage of agents which are in the remain state during the simulation and the bottom right plot the kinetic energy mean at each iteration. This shows that in a phase where most agents no longer change positions and gradually lose their kinetic energy over time due to holding less monetary value the internal energy will descend to zero.  \label{fig:InjectionResultsGini}}
\end{figure}

Figure~\ref{fig:InjectionResultsGini} has analogous plots to Figure\ref{fig:InitialResultsGini} showing the effect of monetary injections on the system. The top left plot shows that the equality of the potential energy is disrupted since the agents which still seek identity satisfaction will move while they contain sufficient funds. The top right shows how the injections affect the inequality of the kinetic energies. The kinetic energy inequality spikes right before almost all the agents have the lowest possible monetary holding bin value so that only a single agent remains with any monetary holding (maximum monetary and kinetic energy inequality). The bottom left shows the agent remain percentage over the simulation trajectory. The bottom right shows the spikes in the model kinetic energies due to the monetary injections at every 200 iterations and how it decreases as the agents hold less monetary values. It can therefore be seen that a system which lacks an injection of energy (monetary values here) degenerates towards fixed point of the system. 

Figure~\ref{fig:InjectionResults} and Figure~\ref{fig:InjectionResultsGini} displays a key outcome of these simulations is that the behavior of the entropy calculated from the energy states of the system is similar to the entropy trace as it would be defined by the canonical entropy formulation from the microstates of the system \cite{mantzaris2018examining}. From the interpretation of the entropy values it can be understood that as the entropy decreases that the number of microstates which occupy the state of the system decreases as well and this provides a different progression metric than other Schelling trajectory methods based upon cluster densities \cite{rogers2011unified}. This does show that the rule set of the Schelling model is not performing randomize state allocations which would result in the state oscillating around the largest microstate mode. 

A key contribution of this paper is the use of the energy states based on the kinetic and the potential energy for the definition of the entropy. In thermodynamics, the entropy is defined on the energy states of the atoms. It is the aim of this work to establish a statistical/informational description of a system that will not only describe its agent behaviors directly, but also be compatible with thermodynamics. This is being done because by utilising this derivation of the entropy the rest of the thermodynamical values such as the energy or temperature will be able to be defined following this work, and allow a model to draw from the scientific knowledge of thermodynamics.
In the canonical derivation of the entropy, since it is defined directly based on the location microstates, the probabilities calculated would not be able to be utilised for a calculation of an internal energy even if the energy state was subsequently defined. As such, any definition of macroscopic energy values in such a system would not be self consistent with the entropy of the system and no conclusions could be made. Only by defining these microscopical energy states of the kinetic and potential energy of the system, can a definition of the macroscopic energy value be developed. Thus the fact that the behavior of the entropy of the system (calculated upon the energy microstates) is behaving similarly to the canonical entropy assures that this definition hold true and can be utilised further in the future in other socio physics models.

\section{Discussion}

This work presents a modified Schelling model which includes a monetary dynamic. In the paper it is referred to as the Monetary Schelling model as agents require a certain amount of monetary value to change grid position and that a cost is incurred for each movement. A certain amount of monetary value is also removed at each iteration so that over time agents have their monetary store moving to zero. The overarching goal of the paper is to provide an incremental advancement on the goals of the field of Socio Physics, the agents have a potential energy and kinetic energy defined based on their state at each iteration. Using the quantities of the energies the entropy of the model at each time point as well as the internal energy is defined. This provides a novel approach to calculating the entropy for a computational model of a social process. 

The results showcase how there are agents which cease to explore grid positions for identity satisfaction when an insufficient amount of monetary store values exist and that monetary injections are required for further explorations of new positions. The injections demonstrate that the system will move towards a fixed stable state without monetary injections as the kinetic energy (based on monetary holdings of the agents) is lost to a process of monetary decrease (akin to a loss in thermal radiation). 
It can be seen how the formulations of the kinetic and potential energy for agents can be defined and be incorporated into the modified Schelling model where the entropy is based upon those energy state values. 
In terms of policy the applicability of such a modeling approach can inspire the computation of energy within groups that have power brokers between organizations with competitive alignments \cite{angst2022information}.

\bibliographystyle{plain}
\bibliography{references}

\end{document}